# SOLVING THE INVERSE KNUDSEN PROBLEM: GAS DIFFUSION IN RANDOM FIBROUS MEDIA

Wojciech Szmyt[a,b,c,d,1]*, Carlos Guerra-Nunez[e], Clemens Dransfeld[f] and Ivo Utke[e]

[a] Institute of Polymer Engineering, FHNW University of Applied Sciences and Arts Northwestern Switzerland, Klosterzelgstrasse 2, CH 5210 Windisch, Switzerland
[b] Laboratory for Micro- and Nanotechnology, Paul Scherrer Institute, CH 5232 Villigen PSI, Switzerland
[c] Department of Physics, University of Basel, Klingelbergstrasse 82, 4056 Basel, Switzerland
[d] Swiss Nanoscience Institute, University of Basel, Klingelbergstrasse 82, CH 4056 Basel, Switzerland
[e] Mechanics of Materials and Nanostructures Laboratory, EMPA Swiss Federal Laboratories for Material Science and Technology, Feuerwerkerstrasse 39, CH 3602 Thun, Switzerland
[f] Aerospace Manufacturing Technologies, Delft University of Technology, Kluyverweg 1, 2629 HS Delft, The Netherlands

*Corresponding author. Wojciech.Szmyt@empa.ch

**Abstract**

About a century ago, Knudsen derived the groundbreaking theory of gas diffusion through straight pipes and holes, which since then found widespread application in innumerable fields of science and inspired the development of vacuum and related coating technologies, from academic research to numerous industrial sectors. Knudsen's theory can be straightforwardly applied to filter membranes with arrays of extended holes for example, however, for the inverse geometry arrangement, which arises when solid nanowires or fibers are arranged into porous interwoven material (like in carpets or brushes) the derivation of an analytical theory framework was still missing. In this paper, we have identified the specific geometric and thermodynamic parameters that determine the gas diffusion kinetics in arrays of randomly oriented cylinders and provide a set of analytical expressions allowing to comprehensively describe the gas transport in such structures. We confirmed analytical solutions by Monte Carlo simulations. We specify our findings for an atomic layer deposition process, the diffusion of trimethyaluminium molecules into a carbon nanotube array, but highlight the applicability of our derivation for other fields comprising gas diffusion membranes, composite materials, fuel cells and more.

**List of symbols**

Latin

    $d$     average fiber diameter, m

    $D_b$     diffusivity in the bulk gas, cm$^2$s$^{-1}$

    $D_{eff}$     effective diffusivity, cm$^2$s$^{-1}$

    $D_K$     Knudsen diffusivity, cm$^2$s$^{-1}$

---

[1] Present address: Mechanics of Materials and Nanostructures Laboratory, EMPA Swiss Federal Laboratories for Material Science and Technology, Feuerwerkerstrasse 39, CH 3602 Thun, Switzerland

Abbreviations: ALD atomic layer deposition; CNT carbon nanotube; CVD chemical vapor deposition; CVI chemical vapor infiltration; MC Monte Carlo

| Symbol | Description |
|---|---|
| $\widetilde{D}_\text{eff}$ | effective diffusivity reduced, dimensionless |
| $\widetilde{D}_\text{K}$ | Knudsen diffusivity reduced, dimensionless |
| $\widetilde{D}_\text{p}$ | diffusivity in the viscous flow regime reduced, dimensionless |
| $d_\text{m}$ | kinetic diameter of a gas molecule, Å |
| $D_\text{p}$ | diffusivity in the viscous flow regime, cm²s⁻¹ |
| $J$ | gas flux through the membrane, cm⁻²s⁻¹ |
| $k$ | number of spatial dimensions, dimensionless |
| $k_\text{B}$ | Boltzmann's constant, m²kg s⁻²K⁻¹ |
| Kn | Knudsen number, dimensionless |
| $l$ | thickness of the fibrous membrane, mm |
| $L_z$ | axial length of the simulation volume cylinder, μm |
| $l_0$ | average fiber length, mm |
| $n$ | gas concentration, m⁻³ |
| $N_\text{A}$ | Avogadro's constant, mol⁻¹ |
| $p$ | gas pressure, Pa |
| $\text{pdf}_\theta$ | probability density function of the polar angle, rad⁻¹ |
| $\text{pdf}_\varphi$ | probability density function of the azimuthal angle, rad⁻¹ |
| $\text{pdf}_x$ | probability distribution of the molecule flight path length, m⁻¹ |
| $R$ | universal gas constant, m³Pa K⁻¹mol⁻¹ |
| $r$ | distance between fiber axis to the molecule flight trajectory, μm |
| $\vec{r}_0$ | origin point of a generated cylinder, μm |
| $T$ | Temerature, K |
| $t_\text{D}$ | diffusion-driven infill time, s |
| $t_\text{f}$ | mean time of flight of a molecule in the space confined by the fibrous structure, s |
| $v$ | mean absolute thermal velocity of gas molecule, m s⁻¹ |
| $x_0$ | x-coordinate of the origin point of a generated cylinder, μm |
| $y_0$ | y-coordinate of the origin point of a generated cylinder, μm |
| $z_0$ | z-coordinate of the origin point of a generated cylinder, μm |

Greek

| Symbol | Description |
|---|---|
| $\alpha$ | surface area to volume ratio of membrane composed of randomly-oriented cylinders, m⁻¹ |
| $\alpha_0$ | surface area to volume ratio of membrane composed of non-intersecting cylinders, m⁻¹ |
| $\Delta l$ | the total length of the fibers within the considered volume element, m |
| $\delta l$ | length of the fiber axis element crossing the molecule flight path collision cylinder, m |
| $\Delta N$ | number of gas molecules in the volume element considered, dimensionless |
| $\Delta \vec{r}$ | spatial displacement of a single step in the random walk or orientation vector of a generated cylinder, μm |
| $\delta s$ | length of the projection of the fiber axis element crossing the molecule flight path collision cylinder onto the plane perpendicular to the flight path, μm |
| $\Delta S$ | surface area element, m² |
| $\Delta t$ | duration of a single step in a random walk, s |
| $\Delta V$ | element of volume considered, m³ |
| $\Delta \vec{x}$ | molecule flight path vector, μm |
| $\varepsilon$ | porosity of a membrane composed of randomly-oriented cylinders, dimensionless |
| $\varepsilon_0$ | porosity of a membrane composed of non-intersecting cylinders, dimensionless |
| $\Gamma$ | gas impingement rate, m⁻²s⁻¹ |
| $\Gamma_b$ | classical gas impingement rate, m⁻²s⁻¹ |
| $\lambda^2$ | mean square flight distance of a molecule within the fibrous structure, nm² |
| $\lambda_b$ | mean free path in the bulk gas, nm |
| $\lambda_f$ | mean flight path in the space confined by the fibrous structure, nm |
| $\mu$ | molar mass (subscript indicates the species), g mol⁻¹ |
| $\varphi$ | azimuthal angle, rad |
| $\sigma$ | fiber axis length per volume, cm⁻² |
| $\sigma_0$ | areal density of fibers, cm⁻² |
| $\tau$ | tortuosity of the random fibrous structure, dimensionless |
| $\tau_{eff}$ | effective tortuosity, dimensionless |
| $\tau_K$ | Knudsen tortuosity, dimensionless |
| $\tau_p$ | tortuosity of the porous structure, dimensionless |
| $\theta$ | polar angle, rad |

## 1. Introduction

A deep understanding of the gas transport in porous media is fundamental in many fields, spanning from the development of fuel cells [1–3], molecular sieving [4], direct contact membrane distillation [5–7], geologic carbon sequestration to management of nuclear waste, and many more [8]. Moreover, knowledge of the gas transport specifics is particularly important in material science, e.g. for a uniform coating of complex porous structures from gas-phase deposition techniques, such as in atomic layer deposition (ALD) [9–15] chemical vapor deposition (CVD) [16] or chemical vapor infiltration (CVI), relevant to numerous sectors in industry [17]. The mechanisms of gas transport in the porous media include viscous flow, surface diffusion, molecular diffusion and Knudsen diffusion [18]. Depending on the physicochemical parameters of a system considered, some of the mentioned mechanisms can be neglected, while other ones prevail and govern the process. Particularly, the Knudsen diffusion becomes significant in the so-called *molecular regime* of gas transport, when the mean free path of gas molecules exceeds the characteristic dimension of the pores. This is why an accurate description of the Knudsen diffusion - and the transition to bulk diffusion - is of great importance for the case of tightly porous media.

Ever since the seminal work of Martin Knudsen [19–21], it has been established that modelling the gas transport in the molecular regime requires a detailed analysis of the flight paths of individual molecules. In this regime, the paths are unconstrained by the intermolecular collisions, being hindered only by the solid walls of the medium in which the diffusion occurs. Consequently, the geometry of the porous medium plays a key role in the modelling. Often, the molecular gas transport models for particular geometries are based on probabilistic numerical approaches, such as direct Monte-Carlo methods [22–25] or Markov chains [26]. Analytical expressions are available for relatively simple geometries such as cylindrical pores [19,27], rectangular cross-section trenches [28], tortuous capillaries [29] or randomly packed hard spheres [30], to name a few examples. These expressions are often used as approximations for gas diffusion in more complex media. To the best of our knowledge however, there has been no analytical model for the Knudsen diffusion in randomly-oriented fibrous structures.

In a recent work Yazdani *et al*. [10] modelled a case that is close to the one considered here, namely, a gas diffusion in vertically-aligned carbon nanotube (CNT) arrays; with the purpose to uniformly coat the nanostructure with ALD. Therein, a cylindrical pore approximation was employed and an expression for an effective pore diameter was proposed. This approach allowed to grasp the general trends of the diffusion coefficient with respect to the average CNT diameter and areal density. In a subsequent work [31], we introduced a diffusion model derived from basic physical principles and obtained an improved expression for the diffusion coefficient in CNT arrays with ideal vertically-aligned geometry. Here, we culminate in deriving a much more general gas diffusion model for real-world fibrous three-dimensional porous structures. It will allow to greatly expand the range of structures for which the close-form equations for gas transport are available, including electrospun fibers [32], curly CNT arrays [33], gas diffusion layers for polymer electrolyte membranes [1] and more. The previous attempts to describe the gas diffusion within random fibrous media [32,34–36] relied on semi-empirical formulae or probabilistic simulations. Close-form expressions for gas transport parameters in such structures have remained elusive, until now.

The aim of this work was to derive and validate an analytical model for the inverse problem of gas diffusion in random fibrous media. We identify the specific geometrical parameters that determine the gas diffusion kinetics in arrays of randomly oriented cylinders. Expressions for the Knudsen diffusion coefficient and the Knudsen number, which determines the boundary between the Knudsen and bulk diffusion are derived. Moreover, we extend the model to express the effective diffusion coefficient in cases when both Knudsen and bulk diffusion are significant, which is relevant at relatively higher pressures [37]. Additionally, an expression for a gas impingement rate within such structures is given. We validate the analytical model by means of Monte-Carlo simulations.

## 2. Assumptions of the model

To specify the applicability of the model presented in this work, we list and discuss the model assumptions. The model presented in this work accounts for both the Knudsen diffusion and a transition to the bulk gas diffusion. While the bulk gas diffusion is already well-described and understood, initially we describe the diffusion in the purely molecular regime. The continuous extension to transition- and bulk diffusion regime is described in section 3.

In the following derivation, the gas transport within the nanostructure is assumed to occur in a molecular regime, i.e. the mean free path in the bulk gas $\lambda_\mathrm{b}$ at the gas concentrations considered is much longer than the mean flight path of a molecule in the space confined by the walls of the porous structure $\lambda_\mathrm{f}$. It is established for an ideal gas [38], that the $\lambda_\mathrm{b}$ is expressed as

$$\lambda_\mathrm{b} = \frac{k_\mathrm{B} T}{\sqrt{2} p \pi d_\mathrm{m}^2}, \tag{1}$$

where the $k_\mathrm{B}$ is the Boltzmann's constant, $T$ – temperature, $d_\mathrm{m}$ – gas molecule diameter, $p$ – gas pressure. The $\lambda_\mathrm{f}$, referred to in the further part of this work as *mean flight path*, needs to be expressed specifically for the given porous structure geometry and in this work we are establishing it for an array of randomly-oriented cylinders.

Effectively, intermolecular collisions are neglected, only the molecule-wall collisions are accounted for. This assumption comes down to ensure that the Knudsen number Kn, as we defined below, is much higher than 1,

$$\text{Kn} := \lambda_b/\lambda_f \gg 1, \tag{2}$$

while the molecular flow component is stated to become apparent at Kn=1 or greater [39,40]. Notably, in literature, a slip-flow regime is sometimes specified for Kn $\in$ (0.001, 0.1) [1], however in this work we are limiting the division into bulk gas diffusion regime, transition regime and molecular regime, for simplicity and clarity. The Kn is conventionally defined as a ratio of $\lambda_b$ to the inner diameter of the given pipe. It has been established so, because for the cylindrical pores, the $\lambda_f$ is in fact equal to the pore diameter. However, in the inverse Knudsen problem that we are considering, there are no pores *per se*. This is why we are proposing the unambiguous definition of the Knudsen number (2), which is applicable to all kinds of porous structures, including the curly fiber bushes.

We are treating the subsequent flights of a molecule from one wall collision to another as a random walk process, consisting of the following steps:

- Collision of a molecule with a cylinder wall,
- Release of a molecule from the cylinder surface following the Lambert's law of emission,
- Free flight of a molecule from one cylinder wall to another at a speed corresponding to the mean absolute velocity $v$ from the Maxwell-Boltzmann distribution,
- Collision of a molecule with another cylinder wall.

The mean absolute velocity $v$ is expressed as

$$v = \sqrt{\frac{8RT}{\pi\mu}}, \tag{3}$$

where $T$ is temperature, $R$ – universal gas constant, $\mu$ – molar mass of the diffusant.

Due to the strictly random orientations and locations of the cylinders, the angle of emission of a molecule is immediately "forgotten" by the molecule upon release – mathematically, it has no bearing on the molecule flight distance. Notably, this is contrary to the anisotropic case of unidirectional cylinders, where the release angle is in fact the key factor determining the path length between subsequent collisions [31].

Following the Einstein-Smoluchowski model [41,42], the diffusion coefficient for a random walk is expressed as

$$D_K = \frac{\langle|\Delta\vec{r}|^2\rangle}{2k\langle\Delta t\rangle}, \tag{4}$$

where $|\Delta\vec{r}|$ is a single random walk step distance, $\Delta t$ - its duration, whereas $k$ – the number of dimensions in which the diffusion is considered. The triangular brackets represent an expected value of the given random quantity. Given the formula (4) it becomes clear that in order to fundamentally describe the Knudsen gas transport, it is crucial to accurately capture the statistical nature of the molecule scattering, as done for instance by Colson *et al.* for the case of cylindrical nanopores [43].

In general, if the geometry of the porous structure considered is anisotropic, the diffusion needs to be expressed in each independent dimension. For instance, in our previous work [31] where we considered an anisotropic diffusion within arrays of unidirectional cylinders. Therein, the 1-dimensional ($k$=1) diffusion along the direction of fiber axis and the transverse 2-dimensional diffusion ($k$=2) needed to be evaluated separately. In this work however, as the considered geometry of tortuous cylinder arrays is isotropic by nature, the 3-dimentional isotropic diffusion model ($k$=3) is developed.

We define the $t_\mathrm{f}$ as the mean time of flight of a molecule in the space confined by the structure

$$t_\mathrm{f} := \langle \Delta t \rangle \tag{5}$$

and $\lambda^2$ as the mean square flight distance of the molecule in the confined space, or *mean square displacement*

$$\lambda^2 := \langle |\Delta \vec{r}|^2 \rangle, \tag{6}$$

both of which are determined by the geometry of the specific type of the porous structure.

## 3. Generalization towards a transition to a viscous regime at high pressures

Although in the derivations in this work a purely molecular regime of gas transport is considered, the results obtained are straightforwardly generalized to account for a continuous transition to a bulk diffusion regime, where Kn takes values of the order of magnitude of unity or smaller. It is important in cases where the gas pressure is relatively high, e.g. in the spatial ALD [37] or CVD [44]. The effective diffusion coefficient $D_\mathrm{eff}$ can be expressed in terms of the Knudsen diffusivity $D_\mathrm{K}$ and the diffusivity through the porous medium in a purely viscous regime $D_\mathrm{p}$,

$$D_\mathrm{eff} = \frac{1}{\frac{1}{D_\mathrm{K}} + \frac{1}{D_\mathrm{p}}}. \tag{7}$$

In [37] the $D_\mathrm{p}$ has been simply approximated as the bulk gas diffusivity $D_\mathrm{b}$, from the classical gas kinetics [38] being

$$D_\mathrm{b} = \frac{\lambda_\mathrm{b} v}{3}. \tag{8}$$

This approximation however does not account for the fact, that even in a purely viscous gas transport regime the diffusivity of gas through the porous medium is not equivalent to the bulk diffusivity [45]. It is so, because the cross section of gas flow is reduced by the factor of porosity $\varepsilon$ and the concentration gradient is applied over a longer path due to tortuosity $\tau_\mathrm{p}$ of the porous structure. Hence, the expression for the diffusion coefficient in porous structures in viscous regime is

$$D_\mathrm{p} = \frac{\varepsilon}{\tau_\mathrm{p}} D_\mathrm{b}. \tag{9}$$

The geometrical factors $\varepsilon$ and $\tau_\mathrm{p}$ are discussed in more detail in the further part of this work.

### 3.1. Reduced diffusivity

In the context of gas transport in porous media, the diffusivity is often expressed in a reduced form as a ratio between the diffusivity in porous medium to the diffusivity in the bulk gas [46–48]. Particularly, the reduced diffusivity in the viscous regime in becomes

$$\widetilde{D}_\mathrm{p} = \frac{D_\mathrm{p}}{D_\mathrm{b}} = \frac{\varepsilon}{\tau_\mathrm{p}}. \tag{10}$$

We can express the effective diffusivity accounting for both Knudsen and viscous diffusion in the dimensionless terms analogously,

$$\widetilde{D}_{\text{eff}} = \frac{D_{\text{eff}}}{D_{\text{b}}} = \frac{\varepsilon}{\tau_{\text{eff}}}, \tag{11}$$

where the reduced diffusivity is denoted with a $\widetilde{D}_{\text{eff}}$ and an effective tortuosity $\tau_{\text{eff}}$ is implicitly defined. If we additionally define a *Knudsen tortuosity* $\tau_{\text{K}}$ analogously as the $\tau_{\text{p}}$ in equation (9), we obtain

$$D_{\text{K}} = \frac{\varepsilon}{\tau_{\text{K}}} D_{\text{b}}, \qquad \widetilde{D}_{\text{K}} = \frac{\varepsilon}{\tau_{\text{K}}} \tag{12}$$

$$D_{\text{eff}} = \frac{\varepsilon}{\tau_{\text{K}} + \tau_{\text{p}}} D_{\text{b}}, \qquad \widetilde{D}_{\text{eff}} = \frac{\varepsilon}{\tau_{\text{K}} + \tau_{\text{p}}}. \tag{13}$$

The form of equations (11-13) shows, that the tortuosity contributions originating from different factors add up to an effective tortuosity. The physical meaning of $\tau_{\text{K}}$ becomes clear with the completed derivation of the $D_{\text{K}}$ in section 4.5. In short, the Knudsen tortuosity $\tau_{\text{K}}$ can be qualitatively understood as an elongation of the gas molecule trajectory due to multiple molecule-wall collisions, in contrast to the tortuosity $\tau_{\text{p}}$ of the porous structure, which corresponds to the elongation of the pore length relative to the straight line connecting two sides of a membrane.

## 4. Transport of gas between randomly-oriented fibers

Examples of locally randomly oriented fibers are shown in Figure 1. The interwoven CNTs (Figure 1a) were synthesized on an alumina-coated Si wafer by catalytic CVD – the synthesis details available elsewhere [49]. Figure 1b depicts a nanoscale fibrous structure of cellulose aerogel[2], whereas Figure 1c shows mullite microfibers[3]. Similar arrangements can be found in materials like electrospun fiber mats, recycled carbon fiber, *etc*. We are deriving a new set of parameters describing the kinetics of the gas transport within such structures accounting for the specifics of their geometry.

a.  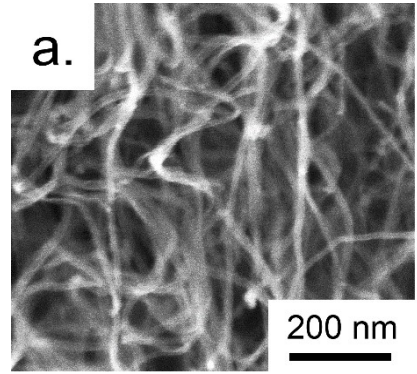 b.

Figure 7c.
from ref. [50]

c.

Figure 5a.
from ref. [51]

Figure 1. Example SEM images of randomly oriented fibrous structures; a) CNTs, b) cellulose aerogel[2], c) mullite fibers[3]

### 4.1. Definition of the structure geometry

Figure 2 schematically illustrates the geometry considered, based on the CNTs as grown by means of catalytic CVD on a planar substrate (Figure 2a) and straight random fibers (Figure 2b). The substrate is shown for illustrative purposes only, while it does not affect the diffusion within the membrane. The fibrous mat is of approximately uniform thickness, $l$. The constituent cylinders are randomly-oriented or tortuous (locally random in orientation), of an average length $l_0$ and diameter $d$.

---

[2] Reprinted from [50] Colloids and Surfaces A: Physicochemical and Engineering Aspects, Vol. 240, Issue 1-3, H. Jin, Y. Nishiyama, M. Wada and S. Kuga, Nanofibrillar cellulose aerogels, Pages 63-67, Copyright 2004, with permission from Elsevier.
[3] Reprinted from [51] Journal of the European Ceramic Society, Vol. 36, Issue 6, J. He, X. Li, D. Su, H. Ji and X. Wang, Ultra-low thermal conductivity and high strength of aerogels/fibrous ceramic composites, Pages 1487-1493, Copyright 2016, with permission from Elsevier.

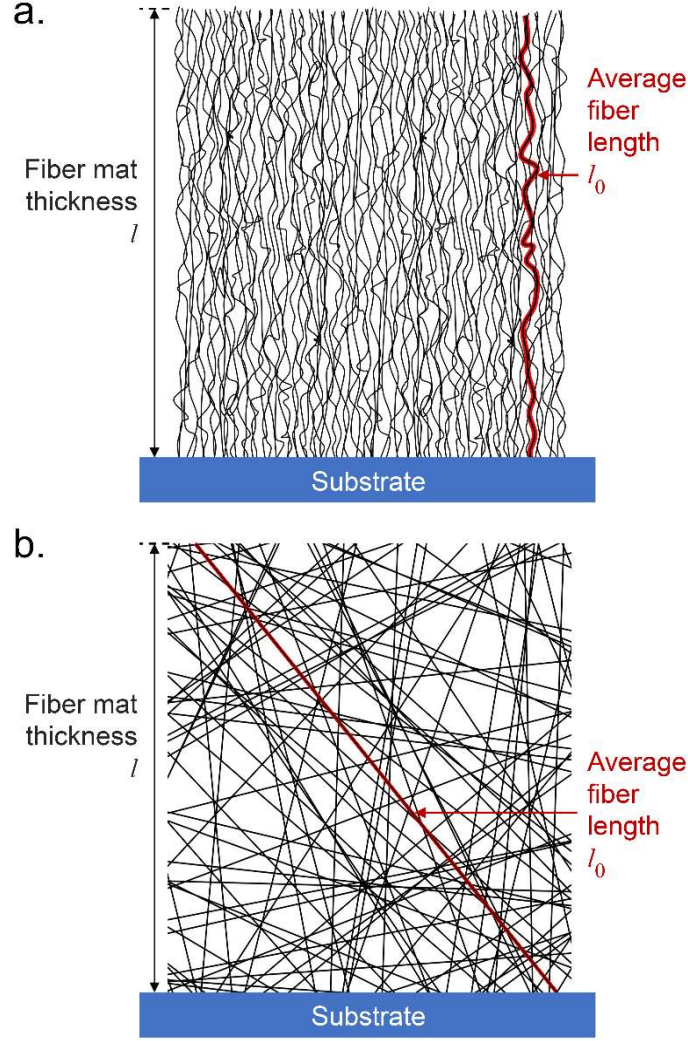

Figure 2. Side-view schematic illustration of mats of randomly-oriented fibers: a) tortuous and b) straight. The average fiber length $l_0$ and fiber mat thickness $l$ are denoted.

In such a case, one can define an *average cylinder axes length per volume σ*

$$\sigma := \frac{\Delta l}{\Delta V} = \frac{\Delta S \sigma_0 l_0}{\Delta V} = \frac{l_0}{l} \sigma_0 = \tau \sigma_0, \qquad (14)$$

where $\Delta V$ is the volume considered, $\Delta S$ – the substrate surface in the representative volume, $\Delta l$ is the total length of fibers in this volume, $\sigma_0$ is the areal density of the cylinders on the substrate, whereas the $\tau := l_0/l$, by one of the common simple definitions, represents *tortuosity* [52]. Consequently, the axes length per volume $\sigma$ also has a meaning of an *effective areal density* of the cylinders on the substrate. For the perfectly isotropic orientations, the $\tau$ equals

$$\tau = 2, \qquad (15)$$

which is shown in the Appendix A. Due to the random and isotropic nature of the defined system, it is safe to assume, that the tortuosity of the pores between the fibers is the same as the tortuosity of the fibers themselves, therefore $\tau$ from equation (15) can be substituted into equation (9) as $\tau_p$.

The other key parameters describing the structure are the surface-area-to-volume ratio $\alpha_0$,

$$\alpha_0 = \sigma \pi d \qquad (16)$$

and the volume fraction of the void or *porosity* $\varepsilon_0$,

$$\varepsilon_0 = 1 - \sigma\frac{\pi d^2}{4}. \tag{17}$$

The simplistic expressions (16,17) hold under an assumption that the fibers are not intersecting or that the intersections between the fibers can be neglected. There are cases however, when the intersections do occur and need to be accounted for. One of such cases is the coating of dense carbon nanotube arrays with thin films, especially if the film thickness is of the order of magnitude of the nanotube diameter or greater. The intersections alter the contribution of individual fibers to the surface area and to the void volume fraction. It is so, because the intersecting cylinder volumes count only once, whereas the intersecting cylinder surfaces do not count at all. The following expressions for the surface area to volume ratio $\alpha$ and porosity $\varepsilon$ do account for the intersecting cylinders:

$$\alpha = \sigma\pi d \exp\left(-\sigma\frac{\pi d^2}{4}\right), \tag{18}$$

$$\varepsilon = \exp\left(-\sigma\frac{\pi d^2}{4}\right). \tag{19}$$

The derivations of (18,19) can be found in the Appendix B. On note is that, equations (18,19) converge asymptotically to equations (16,17), respectively, for small $\sigma\pi d^2/4$ as shown in Figure 4a,b. Moreover, they are applicable for every fiber orientation distribution, such as vertically aligned cylinders, as considered in our earlier work [31], and tortuous fibers, as herein. The $d$ and $\sigma$ are experimentally accessible in a relatively straightforward way, for instance by coupling BET surface area measurements and SEM imaging or directly by means of nano- or microtomography.

Although the expressions (16-19) are not crucial for the further derivations in this work, we are providing them for the benefit of the scientific community, with a view for the experimental evaluation of the geometrical parameters in the forthcoming studies. Nonetheless, while in the analytical derivations we are assuming perfectly random orientations and positions of the cylinders without any hindrance to cylinders crossing, equations (18) and (19) reflect the theoretically considered case.

### 4.2. Regular porous structures vs. inverse porous structures

Equation (18) reveals at which fiber diameter the geometry turns from an inverse-porous to a regular porous structure, e.g. in cases where a growing film infills a nanowire array, aerogel or fiber fabric to a compact nanocomposite in the CVI process. We notice, that it is characteristic for the inverse porous structures that their surface area to volume ratio $\alpha$ increases as the film thickness increases. Examples of such structures include CNT or nanowire arrays, fiber carpets, fibrous aerogels, *et cetera*. The regular porous structures exhibit a decrease in $\alpha$ with the increasing coating thickness, as schematically illustrated in Figure 3. Examples of such porous structures include anodic alumina porous membranes (and others geometrically alike), cracks in solids, microchannels and opals, to name a few. This finding allows us to establish a clear geometrical distinction between the inverse and regular porous structures.

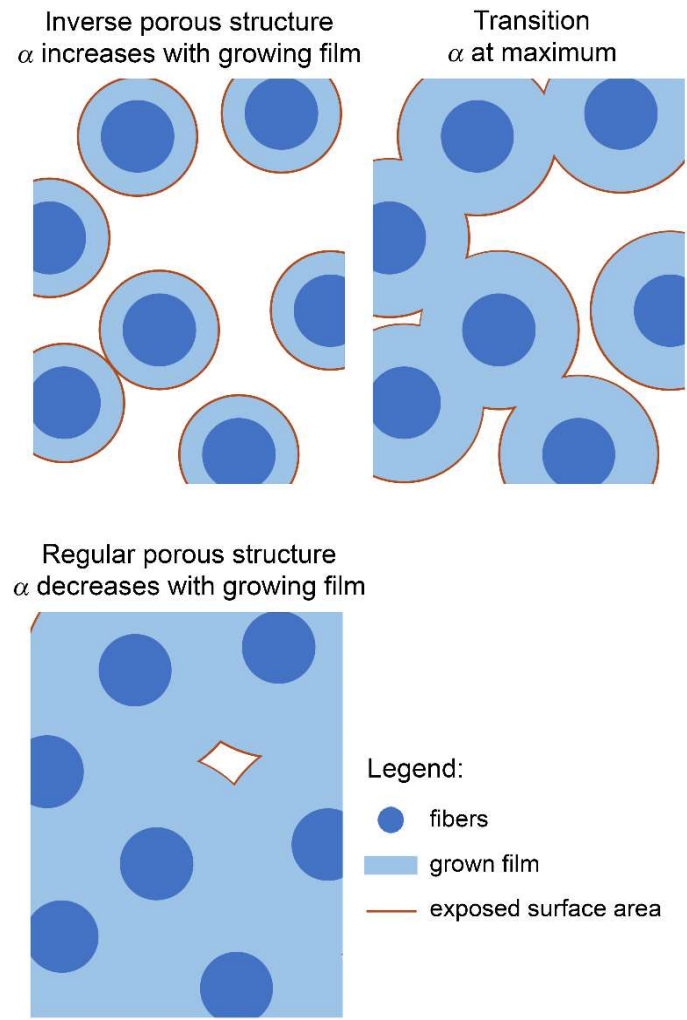

Figure 3. Schematic illustration of the variation in the surface area to volume ratio $\alpha$ with increasing thickness of the coating film infilling the fibrous structure – a transition from an inverse to a regular porous structure.

The expression (18) reflects both the regular- and inverse behaviors and a transition between them, as seen in Figure 4a. The maximum possible value of $\alpha$ for the given $\sigma$ is equal $\sqrt{2\pi\sigma/e}$ and it is reached at the $d=\sqrt{2/(\pi\sigma)}$. This finding is relevant for the applications in which maximizing the surface area to volume ratio of the porous structure is of interest. Interestingly, the $d=\sqrt{2/(\pi\sigma)}$ is also an inflection point of the $\varepsilon$ (19), which means that at this critical cylinder diameter the void volume fraction begins to flatten out to converge asymptotically to 0 at high $d$.

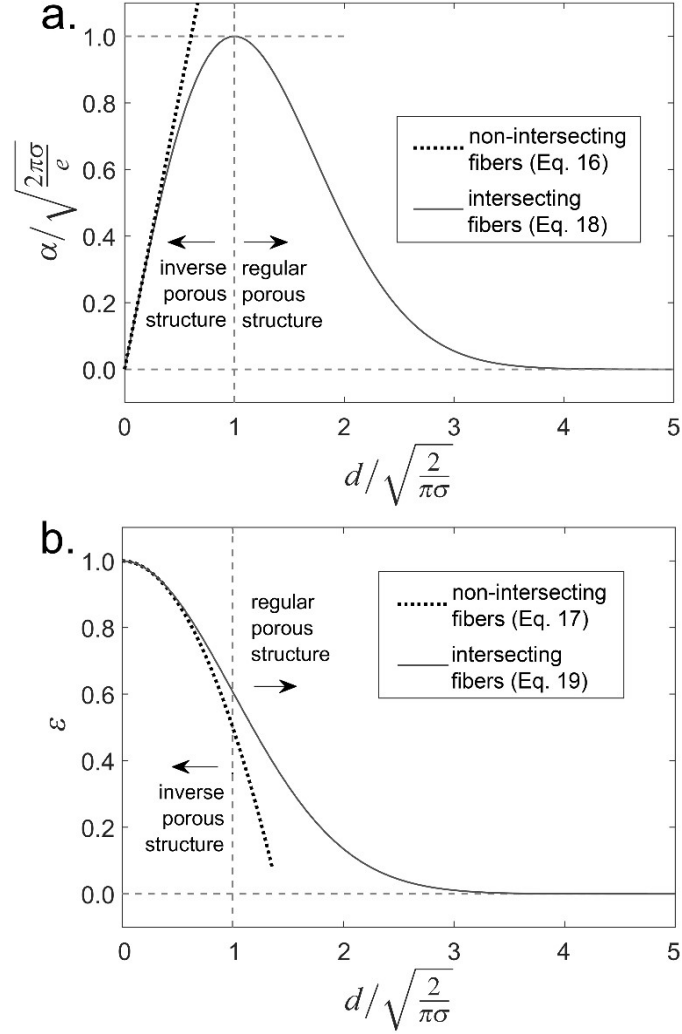

Figure 4. The surface area to volume ratio $\alpha$ (a) and void volume fraction $\varepsilon$ (b) in arrays of the cylinders of diameter $d$. The axes are scaled to the characteristic values for generality. Orthogonal dashed lines were added to guide the eye. Inverse and regular porous structure regimes are indicated. The values of $\varepsilon$ for non-intersecting fibers are plotted up to the physical limit of close-packing of cylinders.

### 4.3. Isotropic fiber orientations – angular distribution

In the derivations, we are assuming that the cylinders are randomly oriented and locally straight, i.e. the radius of curvature is much larger than the individual cylinder diameter. To describe the orientations of the cylinders, we are using the isotropic angle distribution expressed in the spherical coordinate system:

$$\text{pdf}_\varphi(\varphi) = \frac{1}{2\pi}, \qquad \varphi \in [0, 2\pi), \tag{20}$$

$$\text{pdf}_\theta(\theta) = \sin\theta, \qquad \theta \in \left[0, \frac{\pi}{2}\right], \tag{21}$$

where $\varphi$ and $\theta$ represent the azimuthal and polar angle, respectively, whereas the $\text{pdf}_s$ denotes the probability density function of the quantity given in the subscript $s$. Due to the isotropy, the orthogonal reference frame can be defined freely at the convenience of particular derivation.

## 4.4. Mean molecule flight path

Analogously, as in our previous work [31], we derive the mean flight path $\lambda_\text{f}$ and a mean flight time $\tau_\text{f}$ of a molecule in the given porous geometry. For simplicity, we are assuming that the size of the gas molecules is much smaller than the diameter of the cylinders. To account for the finite molecule diameter, one can use the equations obtained here, with the cylinder diameter $d$ increased by $2d_\text{m}$, $d_\text{m}$ being the molecular diameter.

The molecule on a straight path $\Delta\vec{x}$ can only encounter a fiber wall if the fiber axis crosses a cylindrical space of a diameter equal $d$ axially oriented along the molecule flight path $\Delta\vec{x}$, which is illustrated in Figure 5. This cylinder is referred to as *molecule path cylinder*.

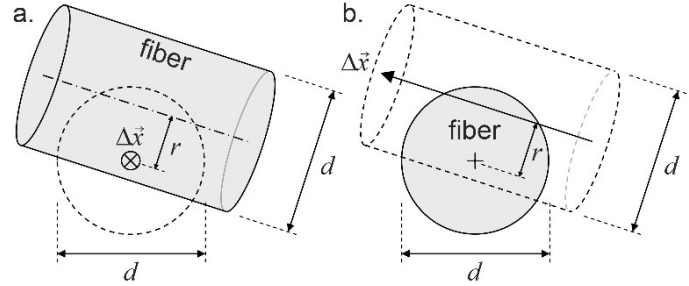

Figure 5. Projected views of the molecule flight path $\Delta\vec{x}$ encountering a fiber of a diameter $d$. The molecule's path cylinder is denoted with dashed lines. Distance $r$ of fiber axis to the axis $\Delta\vec{x}$ needs to be smaller than $d/2$ so that the molecule hits the fiber wall.

The expected length of fiber axes $\Delta l$ within the volume of molecule path cylinder is expressed following the definition (14),

$$\Delta l = \sigma \Delta V = \sigma \frac{\pi d^2}{4} |\Delta\vec{x}|. \tag{22}$$

The $\lambda_\text{f}$ can be defined as the $|\Delta\vec{x}|$, for which the expected number of encountered fibers is 1. In other words, the $|\Delta\vec{x}|$ is equal to $\lambda_\text{f}$ when the $\Delta l$ is equal to the average length of a single fiber axis $\langle \delta l \rangle$ crossing the cylinder, which is expressed mathematically

$$\Delta l = \langle \delta l \rangle \;\Rightarrow\; |\Delta\vec{x}| = \lambda_\text{f}. \tag{23}$$

The geometry that is the basis for the following derivation is illustrated in Figure 6. The distance $r$ of the fiber axis to the molecule path cylinder axis is uniformly distributed,

$$\text{pdf}_r(r) = \frac{2}{d}, \qquad d \in \left[0, \frac{d}{2}\right). \tag{24}$$

It results from the fact that from the perspective along the molecule flight path axis, the fiber presents itself as a two-dimensional stripe-shaped target of a width $d$ (see: Figure 5) and there is no preference to any particular distance $r$. The validity of this assumption is further substantiated by a stochastic simulation in the Appendix E.

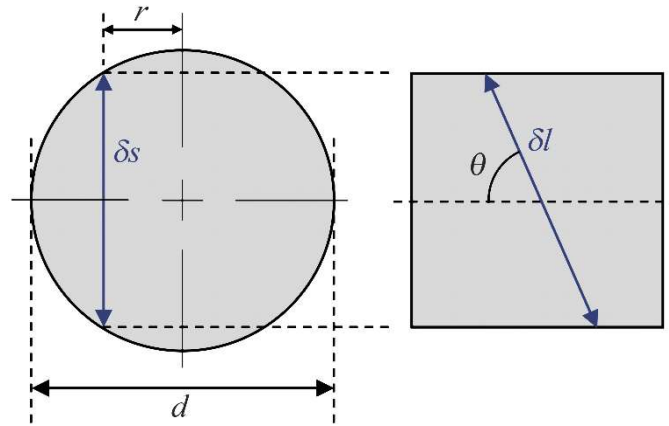

Figure 6. Illustration of the geometry of the fiber axis crossing the molecule path cylinder; projected cross-sectional view along and perpendicular to the molecule path cylinder axis. The $\delta s$ is the length of the element of the fiber axis crossing the molecule flight path cylinder projected onto the plane normal to the flight path. The $\delta l$ is the actual length of the fiber axis element.

In the derivation of $\langle \delta l \rangle$, it is convenient to initially obtain the projection of $\delta l$ onto the plane perpendicular to the $\Delta \vec{x}$, denoted as $\delta s$. Based on the geometry presented in Figure 6, the $\delta s$ is expressed as

$$\delta s = d\sqrt{1 - \left(\frac{2r}{d}\right)^2}. \tag{25}$$

Given the $\delta s$, the angle $\theta$ and the probability distributions (21) and (24), we obtain the $\delta l$

$$\delta l = \frac{\delta s}{\sin \theta}, \tag{26}$$

and its expected value,

$$\langle \delta l \rangle = \int_0^{d/2} \mathrm{pdf}_r(r) \delta s(r) \mathrm{d}r \int_0^{\pi/2} \mathrm{pdf}_\theta(\theta) \frac{1}{\sin \theta} \, \mathrm{d}\theta = \frac{\pi^2 d}{8}. \tag{27}$$

Ultimately, taking equation (22), condition (23), and substituting equation (27), we obtain

$$\lambda_\mathrm{f} = \frac{\pi}{2}\frac{1}{\sigma d} = \frac{\pi^2}{2}\frac{\varepsilon}{\alpha}. \tag{28}$$

In our previous work [31], subsection *Probability distribution of the transverse penetration distance*, we justified that the probability distribution function $\mathrm{pdf}_x(x)$ of the molecule flight path length $x$ follows the Beer-Lambert law,

$$\mathrm{pdf}_x(x) = \frac{1}{\lambda_\mathrm{f}} e^{-\frac{x}{\lambda_\mathrm{f}}}. \tag{29}$$

The same holds for the case considered in this work.

### 4.5. Mean flight time and diffusion coefficient

Knowing the distribution of the molecule flight path and taking the mean absolute velocity of a molecule $v$ as obtained from Maxwell-Boltzmann distribution, we can define the mean flight time

$$t_\mathrm{f} := \langle \Delta t \rangle = \frac{\lambda_\mathrm{f}}{v} = \frac{\pi}{2}\frac{1}{v\sigma d} = \frac{\pi^2}{2}\frac{\varepsilon}{v\alpha}. \tag{30}$$

We express the mean square displacement as

$$\lambda^2 = \langle x^2 \rangle = \int_0^\infty x \, \mathrm{pdf}_x(x) = 2\lambda_\mathrm{f}^2, \tag{31}$$

set $k=3$ in equation (4), as such isotropic diffusion is 3-dimensional, ultimately obtaining the expression for the Knudsen diffusion coefficient of randomly oriented cylinder arrays,

$$D_\mathrm{K} = \frac{2\lambda_\mathrm{f}^2}{2 \cdot 3 t_\mathrm{f}} = \frac{v\lambda_\mathrm{f}}{3} = \frac{\pi}{6}\frac{v}{\sigma d} = \frac{\pi^2}{6}\frac{\varepsilon v}{\alpha}. \tag{32}$$

We can define the Knudsen tortuosity $\tau_\mathrm{K}$ in analogy to the diffusivity in the viscous regime expressed with equation (9),

$$D_\mathrm{K} = \frac{\varepsilon}{\tau_\mathrm{K}} D_\mathrm{b} \Rightarrow \tau_\mathrm{K} := \frac{2}{\pi^2}\lambda_\mathrm{b}\alpha. \tag{33}$$

From equation (33) we can see, that $\tau_\mathrm{K}$ is proportional to the surface area enhancement of a membrane per thickness equal to $\lambda_\mathrm{b}$ and $2/\pi^2$ is a proportionality coefficient. This statement is elaborated on in Appendix C.

### 4.6. Knudsen number

Knowing the $\lambda_\mathrm{f}$, it is straightforward to establish the Knudsen number Kn, following the definition (2),

$$\mathrm{Kn} = \frac{\lambda_\mathrm{b}}{\lambda_\mathrm{f}} = \frac{\dfrac{k_\mathrm{B}T}{\sqrt{2}p\pi d_\mathrm{m}^2}}{\dfrac{\pi}{2}\dfrac{1}{\sigma d}} = \frac{\sqrt{2}\sigma d k_\mathrm{B}T}{\pi^2 p d_\mathrm{m}^2} = \frac{\sqrt{2}}{\pi^3}\frac{\alpha}{\varepsilon}\frac{k_\mathrm{B}T}{p d_\mathrm{m}^2}. \tag{34}$$

### 4.7. Impingement rate

By definition, the impingement rate $\Gamma$ is the number of molecules $\Delta N$ that hit the surface area $\Delta S$ within the time $\Delta t$,

$$\Gamma := \frac{\Delta N}{\Delta S \Delta t}. \tag{35}$$

We set the number of molecules to the average number in the given volume of void within the nanostructure, given the gas concentration $n$ (molecules per unit volume),

$$\Delta N = n\varepsilon \Delta V. \tag{36}$$

The surface area is

$$\Delta S = \alpha \Delta V. \tag{37}$$

The $\Delta t$ is effectively the mean time between collisions, while in this time each molecule should hit the surface once on average,

$$\Delta t = t_f. \tag{38}$$

Putting together equations (35-38), we obtain

$$\Gamma = \frac{n \cdot \varepsilon \Delta V}{\alpha \Delta V \cdot t_f} = \frac{\varepsilon n}{\alpha t_f} = \frac{2nv}{\pi^2} \approx 0.203 \cdot nv, \tag{39}$$

which has the same dependency on the gas concentration $n$ and mean absolute velocity $v$ as the bulk gas impingement rate from the classical gas kinetics $\Gamma_b$,

$$\Gamma_b = \frac{nv}{4} = 0.25 \cdot nv, \tag{40}$$

differing only slightly by a numerical multiplicative factor – approx. 0.203 vs. 0.25. The impingement rate onto cylinder walls in the molecular regime needs to be described with Eq. (39), whereas the impingement rate onto a macroscopic object placed within the fiber bushes (e.g. a substrate or large particles), follows the classical expression for $\Gamma_b$, as expanded on in the Appendix D.

## 5. Comparison of the analytical model with Monte-Carlo simulation results

The key parameter that fundamentally describes the gas transport kinetics is the mean flight path $\lambda_f$, while the other parameters (diffusion coefficient $D$, impingement rate $\Gamma$ and Knudsen number Kn) are intrinsically based on its value. Therefore, we have designed a Monte Carlo (MC) simulation to verify the validity of the derivation of $\lambda_f$ presented in section 4.4. The simulation is based on generation of randomly-oriented cylinders within the specifically chosen simulation volume and measuring the distance of flight of a molecule from the release from a given cylinder wall to the collision with another cylinder.

### 5.1. Simulation algorithm

We have established, that we can limit the simulation volume $\Delta V$ to the molecule path cylinder (see: Appendix E). The geometry for the fiber generation is illustrated in Figure 7. We define the cylinder to be of diameter $d$ equal to the diameter of the fibers and of length $L_z$ equal to 10× the expected mean flight path $\lambda_f$ described with the derived equation (28). Subsequently, the fiber axis lines are generated within the cylinder volume the following way. The origin point $\vec{r}_0 = (x_0, y_0, z_0)$ of each line is randomized, so that

$$\begin{cases} x_0 = r \cos \varphi \\ y_0 = r \sin \varphi \end{cases}, \tag{41}$$

where $r$ is generated by the distribution (24) and $\varphi$ by (20), whereas the $z_0$ is uniformly distributed within $[0, L_z]$. The orientation of the fiber axis is defined by a vector $\Delta \vec{r} = (\Delta x, \Delta y, \Delta z)$, where

$$\begin{cases} \Delta x = -\sin \varphi \sin \theta \\ \Delta y = \cos \varphi \sin \theta \\ \Delta z = \cos \theta \end{cases}. \tag{42}$$

The $\varphi$ is the same as for generating $\vec{r}_0$, and $\theta$ is generated by the distribution (21). A periodic boundary condition is applied to the bases of the simulation volume cylinder, which means that if the line is found to cross any of the bases of a cylinder, it is continued at the other base with the same orientation and origin shifted by $\pm L_z$, the sign depending which base was crossed. Examples of such random fiber generation are visualized in Figure 8, where the cylinders of $d$=20 nm and $\sigma$=$10^{11}$ cm$^{-2}$ were generated within a cylinder of a larger diameter equal 3×$d$ (Figure 8a-c) and 11×$d$ (Figure 8d) and shown only within the diameter 2×$d$ (Figure 8a-c) and 10×$d$ (Figure 8d). In the simulation we generate the cylinders in $x \in (0, 10 \times \lambda_f)$, however the cylinders in Figure 8 are rendered only within $x \in (0, 3 \times \lambda_f)$ for illustrative purposes.

The length of the fiber axis segment crossing the molecule flight cylinder $\delta l$ is evaluated for each generated fiber. The generation of fibers continues until a total length of fiber axes $\Delta l = \sigma \Delta V$ is reached, as required by equation (14).

We define a 1D grid of 0.5 nm pitch along the flight path coincident with the axis $z$. The evaluation of the flight path length is based on finding the value of $z_r$, corresponding to the $z$ coordinate of release of a molecule from the fiber surface, and finding the $z_c$ corresponding to the collision of a molecule with the other fiber. The flight path length $\Delta z$ is evaluated as a difference between $z_c$ and $z_r$. The exact procedure is described in more detail in the Appendix F. The entire process is repeated multiple times for each choice of $\sigma$ and $d$ in order to obtain enough statistics of the flight path lengths.

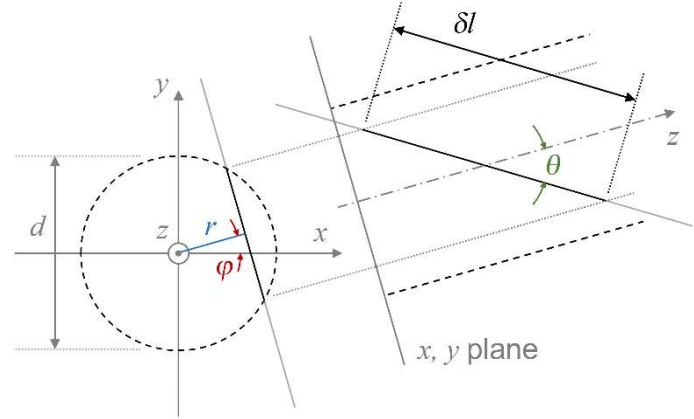

Figure 7. Schematic illustration of the geometry used for the generation of the fibers within the molecule path cylinder presented in two projections. The dashed line represents the molecule path cylinder, the dashed-dotted vector is the collision cylinder axis, the solid line is the axis of the fiber crossing the molecule path cylinder.

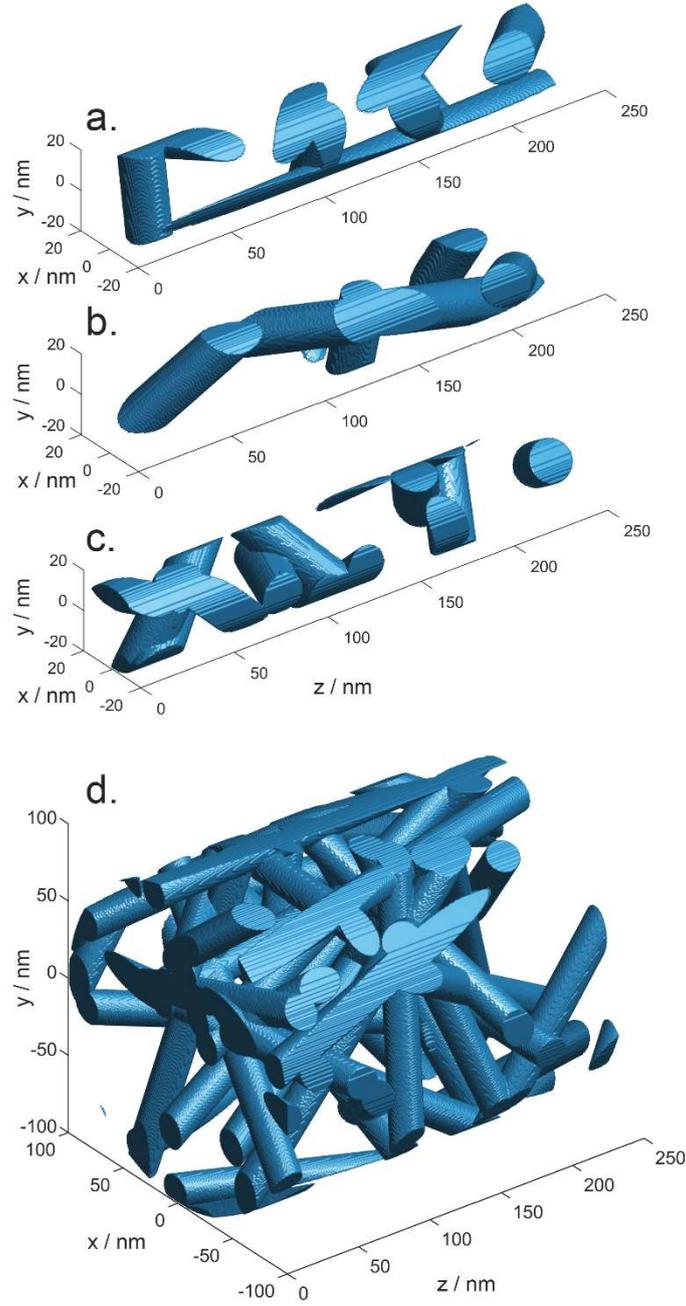

Figure 8. Example of 3D visualizations of the random fiber generation for the diameter $d$=20 nm, axes length per volume $\sigma=10^{11}$ cm$^{-2}$; a-c shown within a cylinder of diameter equal 2×$d$ (a-c) and in a cylinder of diameter 10×$d$ (d).

## 5.2. MC Simulation results compared with analytical model

We evaluated 7 different values of $d$ ranging from 20 to 50 nm at a constant $\sigma$ equal $6\times10^{10}$ cm$^{-2}$ and 5 different values of $\sigma$ ranging from $2\times10^{10}$ to $10\times10^{10}$ cm$^{-2}$ at a constant $d$ equal 30 nm. The number of repetitions was chosen to be $10^6$. Histograms of the set of obtained values of $\Delta z$ have been evaluated at 20 bins between 0 and $5\times\lambda_f$, normalized to the number of counts in the first bin. The histograms were compared to the analytically obtained probability density functions of the mean flight path (29), where the values of the function (29) were proportionally rescaled in a way for the function to return 1 at the location of the first bin. The results are presented in Figure 9.

Page 18

The agreement between the analytically obtained flight path distribution and the one obtained by the stochastic simulation is excellent, which allows us to confirm that the formula for the mean flight path $\lambda_f$ derived in section 4.4 is correct, as well as the $t_f$, $D$, Kn and $\Gamma$, all of which depend on the $\lambda_f$ directly.

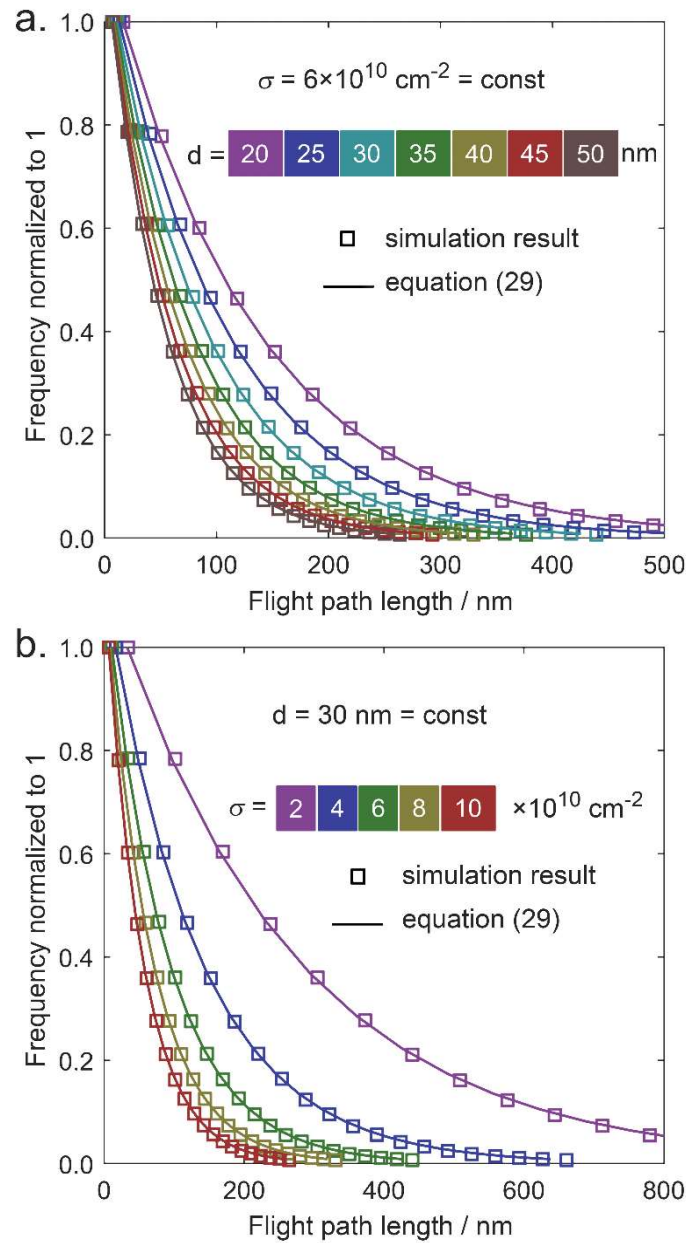

Figure 9. The histogram data of the flight path length distribution compared to the analytically evaluated probability distribution; a) varying fiber diameter $d$ at constant fiber axes length per volume $\sigma$, b) varying $\sigma$ at constant $d$. We are emphasizing, that the solid lines are not a fitting result, but they are directly calculated using equation (29) and normalized to the height of the first bin. The number of MC repetitions $10^6$ ensures that the relative uncertainties of the bin heights are negligible.

## 6. Practical calculations

### 6.1. Example calculation of diffusion coefficient for atomic layer deposition on carbon nanotube mat

In this section, we are evaluating the diffusion coefficient for a model case of atomic layer deposition of $Al_2O_3$ on arrays of CNTs. The diffusing species that we consider here is the trimethylaluminium (TMA) gas precursor–used typically for ALD of $Al_2O_3$ together with $H_2O$, $O_2$ or $O_3$ as the oxygen precursor [53]. We set the temperature to 225 °C (498.15 K) and pressure equal to 1 mbar. At these conditions the TMA exists primarily in a monomeric form, although a significant amount of a dimer might be present as well [54]. For simplicity, we are considering only the monomer of a molar mass equal $\mu_{TMA}$=72.09 g/mol. The diffusion coefficient has been evaluated for a range of values of $\sigma$ from $2\times10^{10}$ to $10\times10^{10}$ cm$^{-2}$ and $d$ from 7 to 30 nm, realistic for the ALD-coated CNTs. The mean absolute velocity is calculated from the Maxwell-Boltzmann distribution. The results are shown in Figure 10a. The diffusivity exhibits a decreasing behavior both with respect to the increasing $d$ and $\sigma$.

The same conditions were used to calculate the effective diffusivity using equation (7) at a range of pressures from $10^0$ to $10^4$ mbar, corresponding to the transition from the molecular regime to the viscous regime of gas flow. As an example, let us consider the value of $d$=20 nm. The results of the calculations described previously are shown in Figure 10b, where the bulk diffusivity and Knudsen diffusivity top limits are indicated as well. Notably, the Knudsen diffusivity does not depend on the pressure, because in this regime the intermolecular collisions are neglected. However, at an increasing pressure, the bulk diffusivity becomes a limiting factor for the diffusion, which comes from the fact, that at higher gas concentrations the intermolecular interactions begin to play a role.

Additionally, we evaluated the characteristic time of diffusion-driven infill (alternatively: evacuation) of the structure with gas $t_D$. It is estimated as a time, for which the diffusion length equals the thickness of the membrane. Consider the CNT mat of a thickness $l$=1 mm. While the diffusion length can be estimated as $(Dt)^{0.5}$, we obtain

$$t_D = \frac{l^2}{D}. \qquad (43)$$

The estimation is shown in the right axis of the graph in Figure 10b. We need to emphasize, that the $t_D$ is not equivalent to the time required for a conformal coating of the CNTs with ALD. For this purpose a scaling law for conformal coating of CNTs needs to be derived accounting for surface reactions, analogously as in [10]. It is however, a good estimate of the time for which a steady state of diffusion is reached within the membrane.

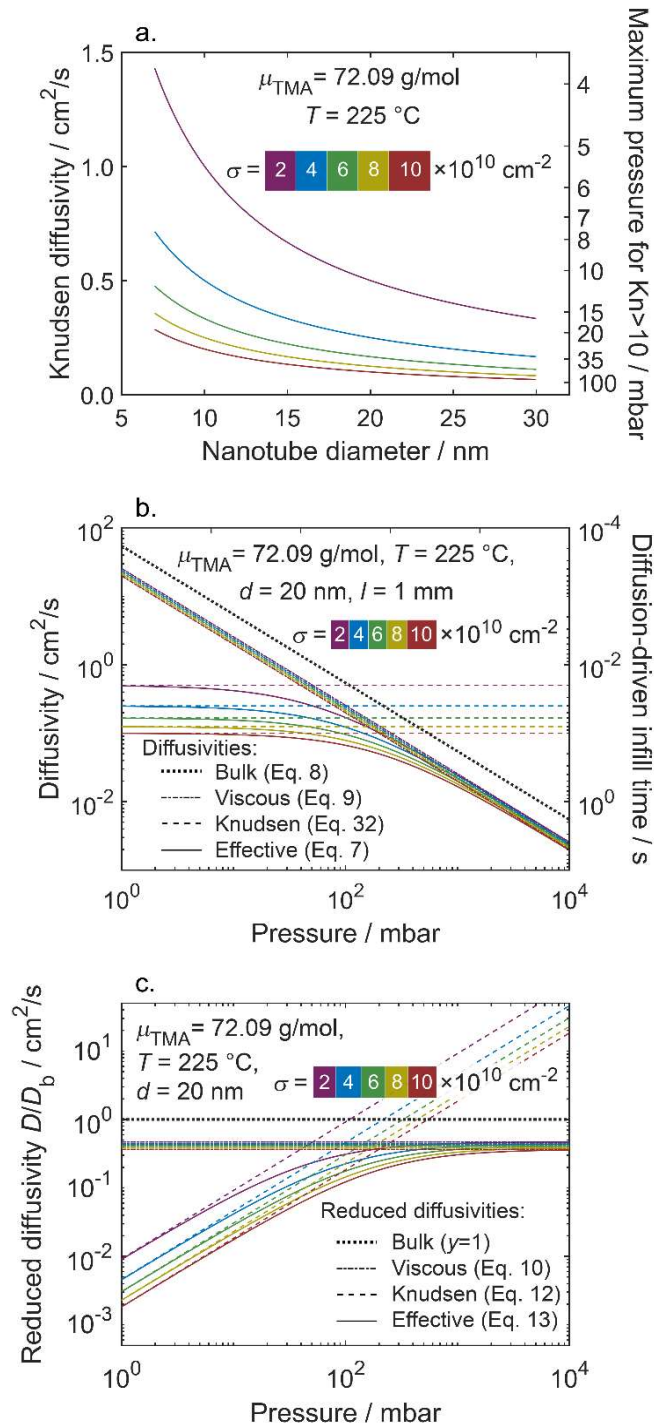

Figure 10. Example calculations of the diffusivity of TMA in a molecular regime (a) and in a transition regime (b,c) in arrays of tortuous carbon nanotubes for given values of nanotube diameter $d$ and fiber axes length per volume $\sigma$. The diffusivity is presented both directly (b), as well as in a reduced form (c), normalized to the bulk diffusivity. The bulk diffusivity is plotted for reference. Additionally, a maximum pressure for the Knudsen regime (Kn>10) is indicated on the right axis of (a) and a diffusion-driven infill time is indicated on the right axis of (b).

### 6.2. Gas flux through a fibrous membrane in Knudsen regime

In this paragraph, we provide an example implementation of our diffusion theory in estimation of fluxes of different gases through a fibrous membrane, specifically in the Knudsen regime. A flat membrane is considered as a simple example, however one needs to keep in mind, that the theory is applicable to more complex geometries as well.

The diffusion-driven gas flux within the membrane $J$ is expressed as [55]

$$J = -D_K \frac{dn}{dz}, \qquad (44)$$

where $n$ is the gas concentration in terms of number of molecules per volume, $z$ – position coordinate across the membrane. In this formulation, the flux $J$ refers to the number of molecules crossing the unit surface area of the membrane per unit time. Let us consider the case, when the concentration of the diffusant on one side of the membrane is $n_D$, whereas beyond the membrane it is negligible or equal 0, which is fulfilled either when the membrane is exposed to vacuum or to another gas, without the diffusant under consideration. The system is schematically illustrated in Figure 11.

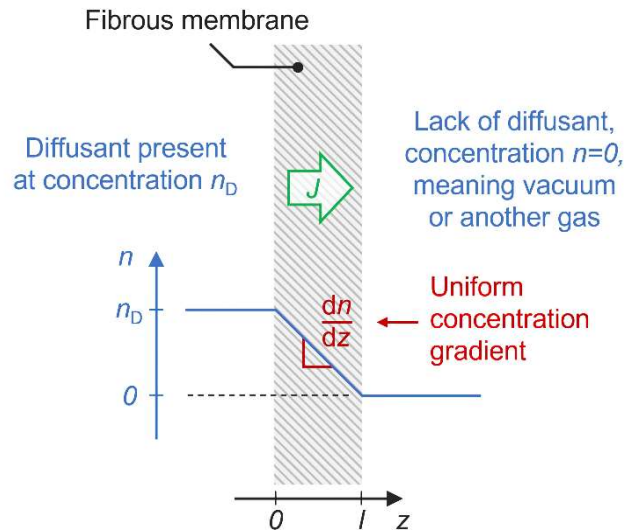

Figure 11. Schematic illustration of the considered system of gas flow through a fibrous membrane. Gas concentration and position coordinate are indicated with the respective axes.

Integrating equation (44) across the coordinate $z$, we obtain

$$J = D_K \frac{n_D}{l}. \qquad (45)$$

Furthermore, we substitute the diffusivity $D_K$ from equation (32), mean absolute thermal velocity of gas molecules $v$ from (3) and express $n_D$ in terms of temperature $T$ and pressure $p_D$, utilizing the ideal gas law. Thus, the expression for the gas flux becomes

$$J = \frac{1}{3}\sqrt{\frac{2\pi}{\mu_D RT}}\frac{1}{l\sigma d} N_A p_D = \frac{1}{3}\sqrt{\frac{2\pi}{\mu_D RT}}\frac{\pi\varepsilon}{\alpha l} N_A p_D. \qquad (46)$$

The form of equation (46) reveals that at the given diffusant pressure $p_D$, the flux is inversely proportional both to the square root of temperature $T$ and to the square root of molecular weight of the diffusant $\mu_D$. Owing to the theory of gas diffusion in random fibrous media developed in this work, the dependence of $J$ on the geometric parameters of the membrane is captured by the dimensionless multiplicative term $(l\sigma d)^{-1}$ or $\pi\varepsilon(\alpha l)^{-1}$, equivalently.

We demonstrate the calculations for gases covering a range of molar masses from 4 to 146 g/mol: helium (He), water ($H_2O$), oxygen ($O_2$), chlorine ($Cl_2$) and sulfur hexafluoride ($SF_6$). The fluxes are estimated for temperatures from 20 to 300 °C (293 to 573 K). We consider a CNT membrane of thickness $l$=1 mm, axes length per volume $\sigma$=6×10$^{10}$ cm$^{-2}$ and CNT diameter $d$=20 nm. In such a membrane, the mean flight path of the gas molecules is calculated as 130 nm, according to equation (28). The diffusant pressure $p_D$=20 mbar is chosen, so that the Knudsen number Kn is greater than 10 for all the cases, ensuring the Knudsen gas transport regime. According to equation (34), the lowest Knudsen number is obtained for the largest kinetic diameter $d_m$ at the lowest temperature $T$ considered. Sulfur hexafluoride has the largest kinetic diameter of all the gases considered $d_m$=550 pm [56], whereas the lowest temperature considered is $T$=293 K. Substituting these values to equation (34) together with $p_D$=20 mbar, we obtain Kn≈11.5, which confirms Knudsen flow conditions throughout the parameter range in the calculations. The results are shown in Figure 12. The term $(l\sigma d)^{-1} = 8.3\times 10^{-5}$ for the above membrane geometry.

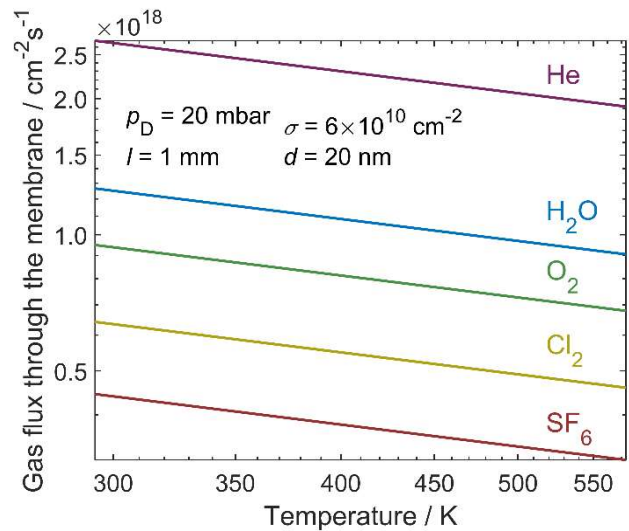

Figure 12. Results of example calculations of gas flux through a fibrous membrane, shown on a double-logarithmic scale plot. In the figure, the physical parameters which were kept constant are indicated: gas pressure $p_D$, membrane thickness $l$, fiber axes length per volume $\sigma$ and fiber diameter $d$. A range of gases has been considered: helium (He), water ($H_2O$), oxygen ($O_2$), chlorine ($Cl_2$) and sulfur hexafluoride ($SF_6$), covering a range of molar masses from 4 to 146 g/mol.

The graph in Figure 12 reveals the dependence of the gas flux on temperature as well as on the molar mass – lighter and cooler gasses diffuse faster at the given pressure, which is characteristic to the Knudsen diffusion and can be utilized for the purpose of gas filtering. The gas fluxes obtained in our calculations are of the order of magnitude of $10^{18}$ molecules per square centimeter per second through the membrane.

## 7. Conclusions and outlook

We have developed a new theoretical framework for the Knudsen diffusion in randomly oriented fibrous media and a transition to bulk gas diffusion regime. The model is derived from the basic physical principles, which gives an advantage over empirical or probabilistic approaches in terms of generality, simplicity and wide applicability. Moreover, the analytical expressions provided allow for straightforward development of scaling laws for various processes involving numerous gas transport processes in various fibrous media. This includes for instance, the prediction of the gas exposure required for a conformal coating of CNTs with ALD, or for a complete infiltration of fibrous structures for composites in CVI, estimation of the growth rate of CNTs synthesized by CVD in case of significant diffusion limitation of the growth, or optimizing the properties of fibrous membranes for the desired gas transport performance in given conditions, among other examples. We have provided a comprehensive set of analytical expressions for the gas diffusion parameters: Knudsen number (34), Knudsen diffusivity (32), effective diffusivity accounting for transition to bulk diffusion (7), mean flight path confined by the structure (28), mean time between subsequent molecule-wall collisions (30) and the impingement rate (39). The set of geometrical parameters determining the Knudsen diffusion is narrowed down to the mean fiber diameter and the fiber length per unit volume, both of which are accessible experimentally in a straightforward way, e.g. by tomography or by coupling microscopy and BET surface area measurements. The present model can serve as a base to generalize the description of gas transport within fibers that are quasi-unidirectional, conceptually between ideally straight (as described in our previous work [31]) and isotropically oriented, like herein. We plan to implement the presented theoretical framework into more complex processes involving e.g. pressure differential, thermal gradient, surface- or gas phase chemical reactions, physisorption, surface diffusion and thermal decomposition. These findings have a substantial impact in various fields of science and technology such as membrane science, composite and nanocomposite technologies, gas phase functionalization, thin film coating by CVD or ALD, CNT synthesis and more.

**Acknowledgements**

We would like to cordially thank the members of the Nanolino group at the Physics Department, University of Basel, Switzerland and the members of the Swiss Nanoscience Institute for fruitful discussions.

**Funding Sources**

This work was supported by the Swiss Nanoscience Institute (SNI PhD project P1402).

**Appendices**

**A. Tortuosity factor of isotropically distributed cylinders**

In section 4.1 we provided the expression for the tortuosity factor of isotropically-oriented cylinders $\tau=2$. In this appendix, we are presenting the derivation that led to this value.

Consider a planar surface $\Delta S$ crossing the fiber carpet at the specific orientation, e.g. parallel to the substrate. The following derivation is based on establishing the number of fibers $\Delta N$ crossing the $\Delta S$, which lets us find the expression for $\sigma_0$ in relation to $\sigma$, and consequently, the tortuosity factor, following the Eq. (14). The angle $\theta$ is measured with respect to the normal to $\Delta S$. We are utilizing the rotational symmetry of the problem with respect to azimuthal rotation by the angle $\varphi$, which simplifies the calculations.

Consider the projection of the surface area $\Delta S$ onto the plane inclined at an angle $\theta$, $\Delta S_\theta$ (see: Figure A.1),

$$\Delta S_\theta = \Delta S \cos \theta. \tag{A.1}$$

The number of fibers that are approximately perpendicular to the $\Delta S_\theta$ comprise of a fraction of all fibers corresponding to the $\mathrm{pdf}_\theta(\theta)$ (Eq. 21), namely

$$dN(\theta)d\theta = \sigma \Delta S_\theta \mathrm{pdf}_\theta(\theta)d\theta = \sigma \Delta S \sin \theta \cos \theta \, d\theta. \tag{A.2}$$

Upon integration over the domain of $\theta$, we obtain

$$\sigma_0 = \frac{\Delta N}{\Delta S} = \sigma \int_0^{\pi/2} \sin \theta \cos \theta \, d\theta = \frac{\sigma}{2}, \tag{A.3}$$

where the $\Delta N$ is the $dN(\theta)d\theta$ integrated over $\theta \in \left[0, \frac{\pi}{2}\right]$. Hence, the tortuosity factor indeed equals

$$\tau = \frac{\sigma}{\sigma_0} = 2. \tag{A.4}$$

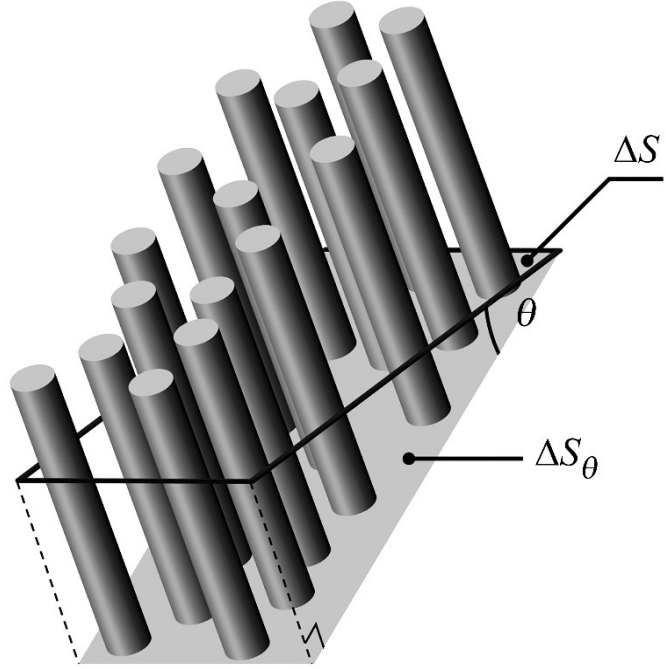

Figure A.1. Illustration of the geometry for the derivation of the tortuosity factor of the randomly oriented fiber carpets. For explanation of symbols, we refer to the main text.

**B. Porosity and surface area to volume ratio within intersecting fiber mats**

In the main part of the work we have presented the equations for $\alpha$ and $\varepsilon$ (18,19) for the case, when the fibers are allowed to intersect. The equations can be used in cases when $\sigma \pi d^2/4 \ll 1$, which means that the average fiber diameter is much smaller than the average spacing between the fibers (effectively: intersections can be neglected), or when films are grown on the fibers, and the film thicknesses are of the same order of magnitude as the fiber diameters or greater, e.g. as grown by means of ALD, CVD or CVI.

It is convenient to begin these derivations from the porosity $\varepsilon$. Consider the structure of fibers of average diameter $d$ and axes length per volume $\sigma$. In the given volume $\Delta V$ there is a total length $\Delta l = \sigma \Delta V$, as per definition (14). The following derivation is constructed as a sequential addition of small fiber lengths $dl$ into the structure, while examining the contribution to the void volume fraction (porosity) $\varepsilon$ of each subsequent addition. The $l \in [0, \Delta l]$ is the independent variable.

If there were no intersections, a single fiber fragment $dl$ would contribute $dl \cdot \pi d^2/4$ to the total fiber volume. However, only the fraction of this contribution crosses the empty space, and this fraction is determined by the current porosity $\varepsilon(l)$. We write

$$d\varepsilon = -dl \cdot \frac{\pi d^2}{4\Delta V} \varepsilon(l) = -d\sigma \frac{\pi d^2}{4} \varepsilon(\sigma), \tag{B.1}$$

where $d\sigma \coloneqq dl/\Delta V$ was defined for convenience. The integration of the above equation should be performed for $\sigma$ from 0 up to its desired value, therefore the collision of symbols between the independent variable and the upper integration limit is not an issue. Given the initial condition

$$\varepsilon(\sigma = 0) = 1, \tag{B.2}$$

the solution of equation (B.1) is

$$\varepsilon = \exp\left(-\sigma \frac{\pi d^2}{4}\right), \tag{B.3}$$

as introduced in section 4.1. We treat the derivation of the $\alpha$ analogously. If there were no fiber intersections, each fiber element $dl$ would contribute $dl \cdot \pi d$ to the surface area, however only a fraction of the added fiber surface crosses the void and this fraction is determined by the already established $\varepsilon$. We write

$$dA_1 \coloneqq dl\pi d \varepsilon = d\sigma \Delta V \pi d \exp\left(-\sigma \frac{\pi d^2}{4}\right), \tag{B.4}$$

Where $dA_1$ is the first contribution of the fiber element to the surface area. There is however also a second - negative - contribution. As the fiber element crosses the existing structure, the already present surface area that is crossed by the fiber is excluded. The fraction of the excluded area is equal to the volume fraction occupied by the fiber element $dl$. We write

$$dA_2 \coloneqq -\frac{dl \frac{\pi d^2}{4}}{\Delta V} A = -d\sigma \frac{\pi d^2}{4} A, \tag{B.5}$$

where $A$ is the current surface area. Setting $dA = dA_1 + dA_2$, we obtain

$$dA = d\sigma \Delta V \pi d \exp\left(-\sigma \frac{\pi d^2}{4}\right) - d\sigma \frac{\pi d^2}{4} A. \tag{B.6}$$

Dividing both sides of the equation by $\Delta V$ and rearranging terms, we transform equation (B.6) to

$$\frac{d\alpha}{d\sigma} = \pi d \exp\left(-\sigma \frac{\pi d^2}{4}\right) - \frac{\pi d^2}{4} \alpha. \tag{B.7}$$

Given the initial condition

$$\alpha(\sigma = 0) = 0, \tag{B.8}$$

we obtain the solution of (B.7)

$$\alpha = \pi \sigma d \exp\left(-\sigma \frac{\pi d^2}{4}\right), \tag{B.9}$$

as introduced in section 4.1. The above derivation is easily generalized to different shapes of randomly added intersecting objects, such as spherical particles, cubes and other irregular shapes, it is however beyond the scope of this work.

## C. Physical meaning of the Knudsen tortuosity

In section 4.5 we derived the Knudsen diffusivity in arrays of random fibrous media. The expression for the Knudsen tortuosity (33) emerged and its physical meaning was outlined. In this appendix we elaborate on it. Let us consider a fibrous membrane of surface area $\Delta S$, thickness $l$ and surface area to volume ratio $\alpha$. The dimensionless surface area enhancement of the membrane can be calculated as

$$\frac{\Delta A}{\Delta S} = \frac{\alpha \Delta V}{\Delta S} = \frac{\alpha \Delta S l}{\Delta S} = \alpha l. \tag{C.1}$$

Hence, $\alpha$ can be understood as a surface area enhancement per unit thickness. Consequently, given equation (33), it becomes clear that $\tau_K$ has a meaning of the surface area enhancement of the membrane of characteristic thickness equal to $2\lambda_b/\pi^2$.

### D. Impingement rate onto a macroscopic object within a fiber carpet

As stated in section 4.7, we must differentiate between the macroscopic impingement rate $\Gamma_c$ and the impingement rate in molecular regime $\Gamma$. Specifically, in this appendix we are showing that the impingement rate within the fiber carpet onto a macroscopic object, such as substrate, is in agreement with the classical gas kinetics. The following derivation is in principle no different than the derivation of the classical bulk gas impingement rate, however we find it necessary to substantiate that the presence of fibers does not change the impingement rate onto macroscopic object. It is of note, that for the simplicity and clarity of the calculations, we treated the molecule movement as if they were always at an average absolute velocity instead of invoking the Maxwell-Boltzmann distribution. It leads however to the same result, while in the following considerations the mean absolute velocity is the determining factor of the kinetics description and the specific profile of the velocity distribution is not relevant.

Consider a very thin slice of the structure adjacent to the macroscopic object surface, illustrated schematically in Figure D.1.

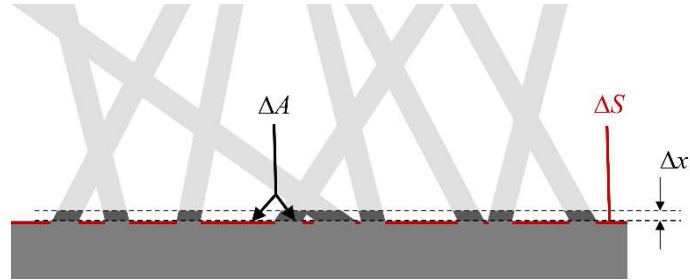

Figure D.1. Illustration supporting the derivation of macroscopic impingement rate.

The slice is so thin, that the molecules present within this slice inevitably either hit the underlying surface or drift away from it; the collision with fiber wall is negligibly likely.

Following the definition (35), we need to establish $\Delta N$, $\Delta t$ and $\Delta S$ in order to obtain $\Gamma_c$. The $\Delta N$ is

$$\Delta N = \frac{n}{2}\Delta V \varepsilon = \frac{n}{2}\Delta A \Delta x \varepsilon. \tag{D.1}$$

The division by 2 is present, because due to the isotropy of the gas flight directions, the movement of only half of all molecules nearby the plane $\Delta A$ is directed at the plane, whereas the other half is drifting away from it. The $\Delta x$ is the thickness of the slice described above. Multiplication by $f_v$ reflects the volume fraction available to the gas molecules.

The surface area of the object available to be reached by the gas molecules is

$$\Delta S = \Delta A \varepsilon, \tag{D.2}$$

Page 27

where $\Delta A$ is the element of the surface area of the object, including the parts obscured by the fibers. Taking only the gas molecules moving towards the object plane ($v_x>0$), their average drift velocity towards the object is

$$\langle v_x \rangle = \int_0^{\pi/2} v \cos\theta\, \text{pdf}_\theta(\theta) d\theta = v \int_0^{\pi/2} \sin\theta \cos\theta\, d\theta = \frac{v}{2}, \quad (D.3)$$

which was calculated using the isotropic distribution of the angle $\theta$, (21). For all the $\Delta N$ molecules within the slice $\Delta x$ to hit the surface, it will take $\Delta t$ time,

$$\Delta t = \frac{\Delta x}{\langle v_x \rangle} = \frac{2\Delta x}{v}. \quad (D.4)$$

Consequently, according to the definition (35), the $\Gamma_c$ is

$$\Gamma_c = \frac{\Delta N}{\Delta S \Delta t} = \frac{\frac{n}{2}\Delta A \Delta x \varepsilon}{\Delta A \varepsilon \frac{2\Delta x}{v}} = \frac{nv}{4}, \quad (D.5)$$

which does remain in agreement with the classical gas kinetics. Exactly the same expression for $\Gamma_c$ holds at the bulk-to-fibers interface, which ensures that the gas concentration within the void of the fibrous structure is in equilibrium with the bulk gas.

### E. Uniform distribution of distance of fiber axes from the flight path

In order to design the MC simulation in an optimal way, it is important to narrow the considered representative volume down to the necessary minimum while avoiding artifacts from boundary effects. In the considered case, the necessary minimum is the molecule path cylinder in which the collisions can occur, as discussed is section 4.4. To accurately simulate the isotropically random orientations and uniformly random positions of fibers in the narrow molecule path cylinder, we have identified two solutions:

a) generating the cylinders in a volume much greater than the volume of interest representing a big part of the porous structure and considering only the fibers that do cross the molecule flight path,
b) generating the cylinders directly in the molecule path cylinder with carefully chosen probability distribution of positions and orientations.

The option b) is much less computationally intensive than a). In order to implement the approach b), we validated the probability distribution of the fiber axis distance from the molecule flight path (24) with the approach a). For this purpose, we considered a molecule path cylinder of 70 nm in diameter and of 1 μm length and we generated straight lines corresponding to fiber axes within the cylinder of 10× larger diameter (700 nm) and of 20× larger length (20 μm), oriented co-axially with the molecule path cylinder. Each line was generated with an origin point uniformly distributed within the large cylinder volume and the orientation vector, following the isotropic distribution (20,21). The sequential line generation was continued until we obtained $10^7$ axes crossing the small molecule path cylinder. Once the generation was complete, we evaluated the histogram of distances of the lines to the axis of the molecule path cylinder and normalized the histogram so that the histogram bar heights correspond to a probability density. Figure E.1 presents the result and a comparison to the distribution (24).

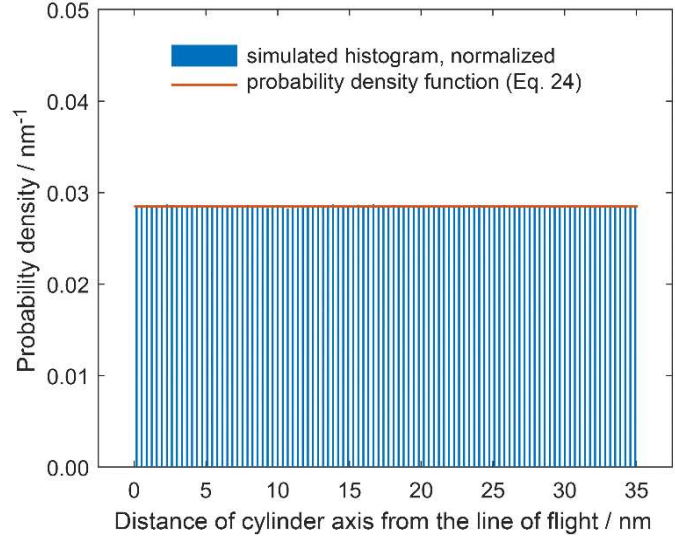

Figure E.1. Normalized histogram of the simulated distances of fibers from the molecule flight path compared to the uniform probability distribution (24).

As we can see in Figure E.1, the distance distribution indeed corresponds to the uniform probability distribution (24), which allows for the optimization of the MC simulation algorithm.

## F. Numerical evaluation of the molecule flight path

We define a 1D grid of 0.5 nm pitch along the flight path coincident with the axis $z$. Each element of the grid is assigned its distance to the closest fiber axis - distances to all lines are evaluated and a minimum value is chosen. We obtain $dist_i$, where $i$ is a natural number indexing the discrete points of the grid and $dist_i$ is the distance value for the given $z_i$. The $z_r$ is identified as the first $z$, for which the distance crosses the value of $d/2$ and is increasing with $z$ at this point, whereas the $z_c$ is found as the first $z$ greater than $z_r$, for which the distance crosses the value $d/2$ but is decreasing. A linear interpolation of the $dist_i$ with respect to real values of $z$ is implemented for an increase in accuracy, as exemplified in Figure F.1. The mean value of all collected values of $\Delta z$ for given $d$ and $\sigma$ is the evaluated mean flight path $\lambda_f$.

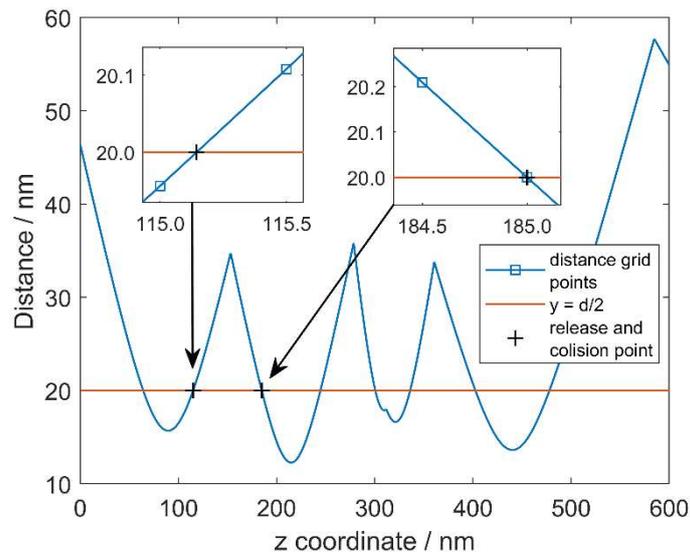

Figure F.1. Example evaluation of the release and collision point based on the linear interpolation of the grid of distance to the nearest fiber axis. For clarity, the square markers of distance grid points are shown only in the insets.